\documentstyle[12pt,aaspp4]{article}
\tighten
\def\etal{et~al.}

\def\OIGS{\:{\rm ergs\:cm^{-2}\:s^{-1}\:\AA^{-1}}}

\def\LA{Ly$\thinspace\alpha$}
\def\LB{Ly$\thinspace\beta$}
\begin{document}

\newcommand{\EXPN}[2]{\mbox{$#1\times 10^{#2}$}}
\newcommand{\EXPU}[3]{\mbox{\rm $#1 \times 10^{#2} \rm\:#3$}}  
			% exponent with units

\title{The Far-Ultraviolet Spectrum and Short Timescale Variability  \\
of AM Herculis \\
from Observations with the Hopkins Ultraviolet Telescope} 

\author{Bradford W. Greeley and William P. Blair}
\affil{Department of Physics \& Astronomy,
The Johns Hopkins University,
Baltimore, MD 21218
greeley@pha.jhu.edu, wpb@pha.jhu.edu}

\author{Knox S. Long}
\affil{Space Telescope Science Institute,
3700 San Martin Drive,
Baltimore, MD 21218
long@stsci.edu}

\and

\author{John C. Raymond}
\affil{Harvard-Smithsonian Center for Astrophysics,
60 Garden Street, Cambridge, MA 02138; raymond@cfa.harvard.edu }

\vskip 1.5 in
\centerline{\small To appear in}
\centerline{\small \em The Astrophysical Journal}

\clearpage
\begin{abstract}

Using the Hopkins Ultraviolet Telescope (HUT), we have obtained 
$850-1850$~\AA\ spectra of  
the magnetic cataclysmic variable star
AM Her in the high state. These observations provide high
time resolution spectra of AM Her in the FUV and sample much of the 
orbital period of the system.  The spectra are not well-modelled in 
terms of simple white dwarf (WD) atmospheres, especially at wavelengths
shortward of \LA.  The continuum flux changes by a factor of 2 near the
Lyman limit as a function of orbital phase; the peak fluxes are observed
near magnetic phase 0.6 when the accreting pole of the WD is most clearly
visible.  The spectrum of the hotspot can be modelled in terms of a
$100\thinspace000$ K WD atmosphere covering $2\%$ of the WD surface.
The high time resolution of the HUT data allows
an analysis of the short term variability and shows the UV luminosity
to change by as much as 50\% on timescales as short as 10 s. 
This rapid variability is shown to be inconsistent with the clumpy
accretion model proposed to account for the soft X-ray excess in polars.
We see an increase
in narrow line emission during these flares when the heated face of the
secondary is in view. The \ion{He}{2} narrow
line flux is partially eclipsed at secondary conjunction, 
implying that the inclination of
the system is greater than $45\arcdeg$.
We also present results from models of the heated face of the
secondary. These models show that reprocessing on the
face of the secondary star of X-ray/EUV 
emission from the accretion region near the WD can account for the
intensities and kinematics of most of the narrow line components observed.

\end{abstract}


\section{Introduction}
Over two decades ago, Tapia (1977) announced the 
discovery of strong temporally varying
circular polarization in the V and I bands in AM Her, and suggested
that it was the optical counterpart of the X-ray source 3U 1809+50
discovered earlier at 2-6 keV by Giacconni \etal\ (1974), and in soft
X-rays (0.15-0.28 keV) by Hearn, Richardson, \& Clark (1976). Tapia
found a maximum linear polarization in the V and I bands of about
5\% and circular polarization in the V band ranging from $+3$\% to
$-8$\%, which he identified as the result of cyclotron emission.
Chanmugam \& Wagner (1977) offered a basic model of this
``Remarkable System'' (as their paper was titled),
consisting of a magnetic WD accreting matter from a red dwarf.
The magnetic field of the WD is strong enough
to prevent formation of an accretion disk.
The rotation of the WD is synchronized with the rotation
of the system, so the accretion stream follows the magnetic field
lines to the surface of the WD. 
Since its discovery, AM Her has become the prototype for a class of 
more than 50 objects known as {\em polars} 
(see Beuermann \& Burwitz 1995, Warner 1995,
and Cropper 1990 for reviews).  

The AM Her system parameters adopted in this paper are summarized in 
Table~\ref{tbl:param}.
With an inclination angle of $35\arcdeg$, the system does not eclipse
in the optical either at secondary conjunction or at magnetic phase
zero (defined as the point at which the 
linear polarization in the V band is maximum).
The luminosity and spectrum vary with orbital phase.
Greenstein \etal\ (1977) discovered that the \ion{He}{1} and \ion{He}{2} lines
in their optical spectra were comprised of broad ($\sim 1000$ km s$^{-1}$)
and narrow components ($\sim 90$ km s$^{-1}$), 
and these shifted in wavelength over the period, almost in antiphase with each
other. They identified the narrow component with the heated face of the 
secondary star and the broad component with the material near the WD.
The velocity curves indicate a phase difference between components of 0.37, 
with the maximum blueshift of the broad component occuring at magnetic
phase 0.03 and the maximum blueshift of the narrow component at magnetic
phase 0.40.

In this paper we will refer to the continuum orbital maximum and minimum 
in our data as the bright and faint phases.  AM Her exhibits a range of 
accretion states but is most often found in a high 
accretion state, with occasional low states lasting on the order of a hundred
days (Feigelson, Dexter, \& Liller 1978).
As material falls toward the WD, a shock forms, producing 
hard X-ray bremsstrahlung emission. It was proposed (Lamb \& Masters 1979,
King \& Lasota 1979) that the cyclotron and bremsstrahlung emission 
heat the atmosphere of the WD, which then radiates the soft X-ray blackbody
component. This cannot be the whole story
for many polars, including AM Her, as the total luminosity of the
blackbody component exceeds the sum of the cyclotron and bremsstrahlung 
components (Rothschild \etal\ 1981; Ramsay \etal\ 1994). 
Kuijpers \& Pringle (1982) suggested that instead
of all of the material encountering a shock, denser clumps in the stream
remain intact and bury themselves in the WD photosphere, heating the 
atmosphere directly. 
This soft X-ray radiation from the heated photosphere 
is then available to heat other parts
of the system including the upper atmosphere of the secondary star and the
accretion stream, where it is reprocessed to UV and optical wavelengths.
Observations of this reprocessed emission offer insight to both the physical
conditions at those sites and the accretion and energy conversion processes 
in the system.   

UV observations with IUE have been important in the development of models
of AM Her. Observations of the UV continuum by Raymond \etal\ (1979),
for example, placed  
important constraints on the nature of the soft X-ray spectrum. Although 
the relatively short 3 hour period of AM Her made obtaining phase resolved
spectra with IUE difficult, coarse light curves were constructed by
Heise \& Verbunt (1988).  
The continuum 
longward of 1200 \AA\ in IUE spectra has been reasonably well fit with a 
$\sim$ 20\thinspace000 K WD atmosphere plus a hotspot component 
for low and high states
(G\"ansicke \etal\ 1995; Silber \etal\ 1996). Studies of the UV lines have
been difficult with IUE, however, because the broad and narrow components
are not well-resolved in the IUE low resolution mode. Phase-resolved
high resolution spectra have been attempted (Raymond \etal\ 1979), but
suffer from low detected signal. Only recently have 
new instruments allowed detailed study of the UV line components with
phase. High resolution sub-\LA\ observations with ORFEUS 
(Raymond \etal\ 1995; Mauche \& Raymond 1998)
clearly show the two component structure in the \ion{O}{6} doublet, 
with the components' wavelength dependence 
as a function of phase similar to that observed for lines in 
the optical. 

In this paper, we present observations of AM Her obtained with the Hopkins
Ultraviolet Telescope (HUT) during the Astro-2 space shuttle mission
in 1995 March. These observations constitute a significant UV data 
set, in terms of wavelength range, phase coverage, and temporal
and spectral resolution, for this prototypical magnetic cataclysmic variable. 
In particular, these observations can be used to study the continuum and
emission line variability on shorter timescales than studied previously.

\section{Observations}

HUT was operated aboard
the space shuttle {\em Endeavour} for 14 days in 1995 March as part of
the Astro-2 mission (STS-67).  The HUT instrument consists 
of a 0.9 m f/2 mirror and a prime focus spectrograph with a first order 
spectral range of $820-1840$ \AA\ and $2-4$ \AA\ resolution.  Davidsen 
\etal\ (1992) describe the instrument in detail as flown on the Astro-1
mission.  For Astro-2, several improvements were made to the telescope
which increased the effective area considerably (Kruk \etal\ 1995).
HUT is a photon-counting instrument and a cumulative histogram is 
downlinked every 2 seconds, enabling the data to be binned into any multiple
of 2 s. The data are then reduced using a calibration pipeline,
developed as an 
IRAF\footnotemark
\footnotetext{IRAF is distributed by the National Optical Astronomy
Observatories, which is operated by the Association of Universities for
Research in Astronomy, Inc.\ (AURA) under cooperative agreement with the
National Science Foundation.}
package specifically for HUT data. The pipeline tasks correct for 
detector pulse-persistence effects, dark count, and scattered light, and
perform wavelength and time-dependent flux calibrations.  The
absolute flux calibration is based on inflight DA WD star observations 
and is generally accurate to $\sim 5\%$ (Kruk \etal\ 1997).

AM Her was observed with HUT three times 
during the Astro-2 mission. The first two
observations were made on consecutive shuttle orbits on 9 March starting
at UT 21:32:02 and 23:10:32, and were part of the same AM Her orbital period.
The last observation was made on
14 March starting at UT 02:42:45, 32 orbital periods later. 
The observations covered magnetic phases
$0.968-0.260$, $0.499-0.662$, $0.667-0.809$, respectively, of the 
186 m orbital period (using the ephemeris given in Heise \& Verbunt 1988).
AAVSO observations indicate that AM Her was in its high state at $V\approx13$.
Post-flight analysis of stored video frames from the HUT slit-jaw camera
indicated that
AM Her was well centered within the 20\arcsec\ slit and that
slit losses were
negligible throughout all of the observations. We also confirmed this
directly during the third observation by changing the aperture from 
20\arcsec\ to 32\arcsec\ while observing AM Her. The 
average source count rate did not change.

\begin{figure}
\plotfiddle{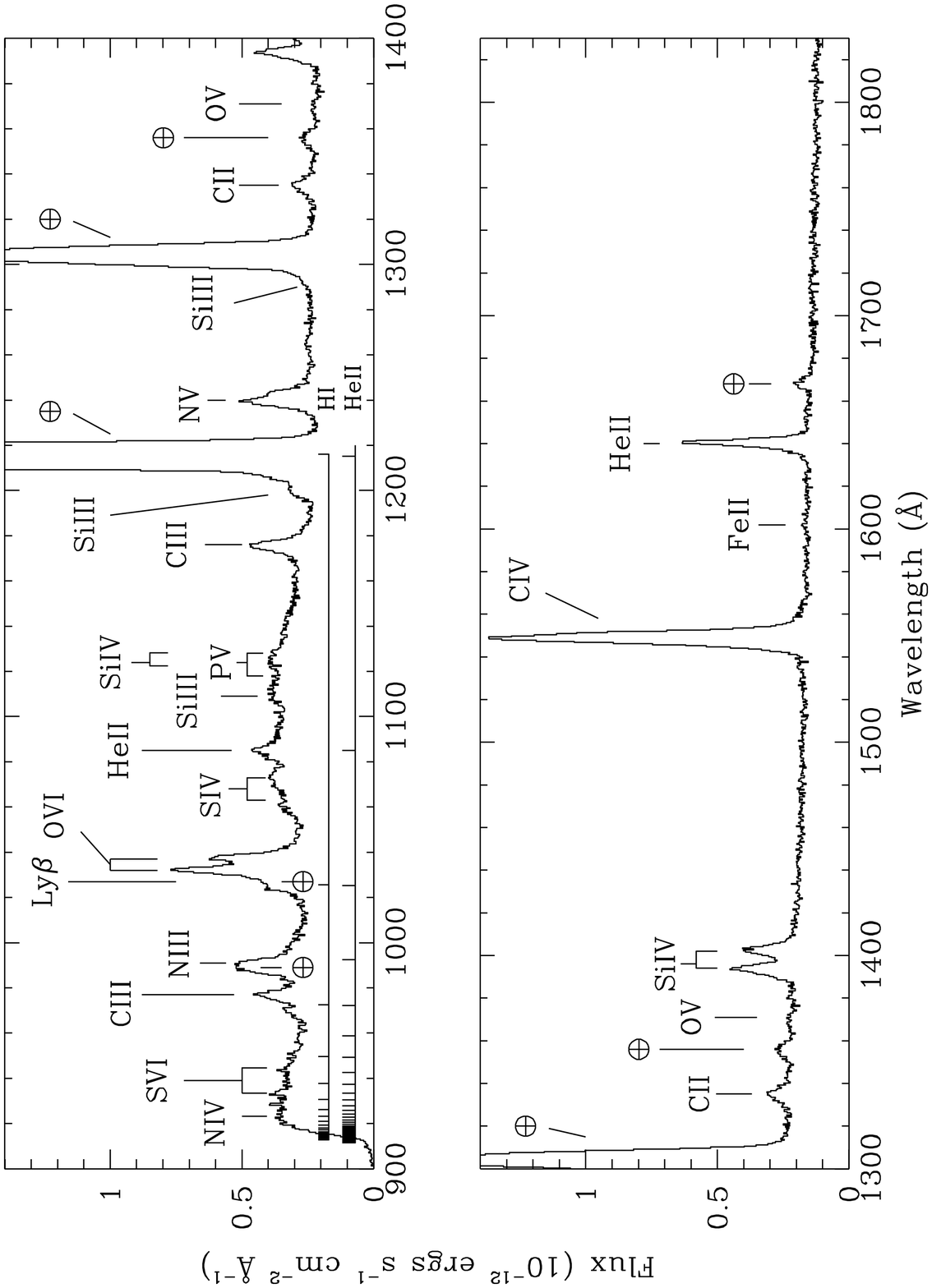}{6.5in}{270}{65}{65}{-254}{450}
\caption{\label{fig:amspec} The summed HUT spectrum of AM Her
exhibits a large range of ion species. Note the line blends from 
$1050-1170$ and $912-960$ \AA. The earth symbol denotes positions of
features expected from terrestrial airglow.
}
\end{figure}

The summed HUT spectrum, covering all 
three observations, is shown in Figure~\ref{fig:amspec}.
The spectrum is characterized by a strong blue continuum and
emission lines from a broad range of ion
species. Probable identifications are indicated
in the figure for some of the more important lines.
The hydrogen Lyman series and \ion{He}{2} Balmer series
line positions are also plotted for reference.  
One or both of these series may contribute
to the apparent continuum rise near the Lyman limit. Another interesting
feature is the excess flux in the $1050-1170$ \AA\ range. This excess may
be a blend of a large number of lines, including those listed in the
figure and possibly \ion{Fe}{2} and \ion{Fe}{3}.

Although this total spectrum shows the overall spectral characteristics in 
exquisite detail, the short time sampling and digital nature of the data 
allow us to investigate the time variability of various spectral components
in detail. In the following sections, we investigate the continuum and
line variations with orbital phase and then discuss the more rapid 
variability seen in these data. 

\section{Analysis}
\subsection{Continuum Variations}

\begin{figure}
\plotfiddle{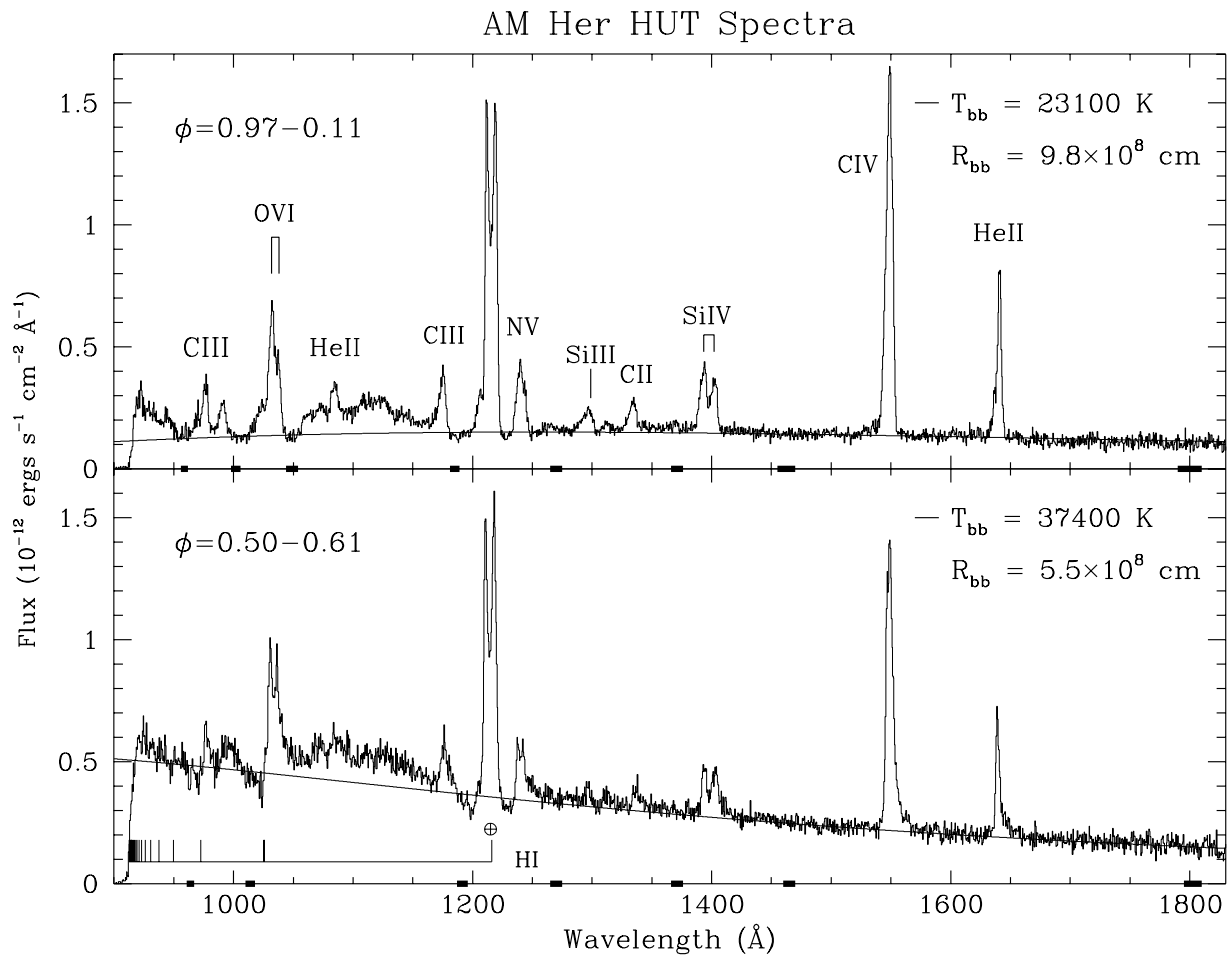}{6.0in}{270}{65}{65}{-254}{450}
\caption{\label{fig:plottwocont} 
The HUT spectrum for minimum (upper panel) and maximum 
(lower panel) light. The most striking difference is the sharp 
rise in the continuum at short wavelengths. Blackbody fits to the barred regions
on the x-axis give continuum temperatures of 23\thinspace100 K and 
37\thinspace400 K
for minimum and maximum, respectively. Also note the change in the line
profiles. Clearly the lines are comprised of narrow and broad components,
which shift with phase. These spectra are from the orbital night portion
of the data and do not contain significant airglow emission other than
\LA.
}
\end{figure}

The spectra for phases $0.97-0.11$ and $0.50-0.61$,
are plotted in Figure~\ref{fig:plottwocont}.
These spectra correspond to the orbital night portions of the first and
second observations and thus do not suffer 
from terrestrial airglow emission other than \LA.
These phases correspond
to the faint and bright phases observed in 
the $75-120$ \AA\ band with EUVE (Paerels \etal\ 1996). 
The mean level of the continuum changed by more than a factor of two at
short wavelengths between the first two observations.
Because the HUT data are described by photon statistics, we can use model
fitting procedures such as those employed by the ``SPECFIT'' task
(Kriss 1994, a task included in the STSDAS IRAF package),
which minimize $\chi^2$ to determine the best fit between model and data.
We fitted blackbodies to the continua using the regions 
denoted by the bars on the wavelength axis of Figure~\ref{fig:plottwocont}.
The temperature of the faint phase fit falls in between the value
of 20\thinspace000 K 
determined with IUE for AM Her in the low state (G\"ansicke \etal\ 1995,
Silber \etal\ 1996) for fits to WD atmospheres and 
$\sim29\thinspace000$ K 
derived from WD fits to the IUE high state faint phase data
(G\"ansicke \etal\ 1995).
The radius of the emission region it implies, 
$9.8\times10^{8}$ cm,
is of order the size of the WD. 
Note that for the bright phase,
the spectrum dips below the blackbody fit near the wavelengths of
Hydrogen \LB, $\gamma, \delta,$ and $\epsilon$.

\begin{figure}
\plotfiddle{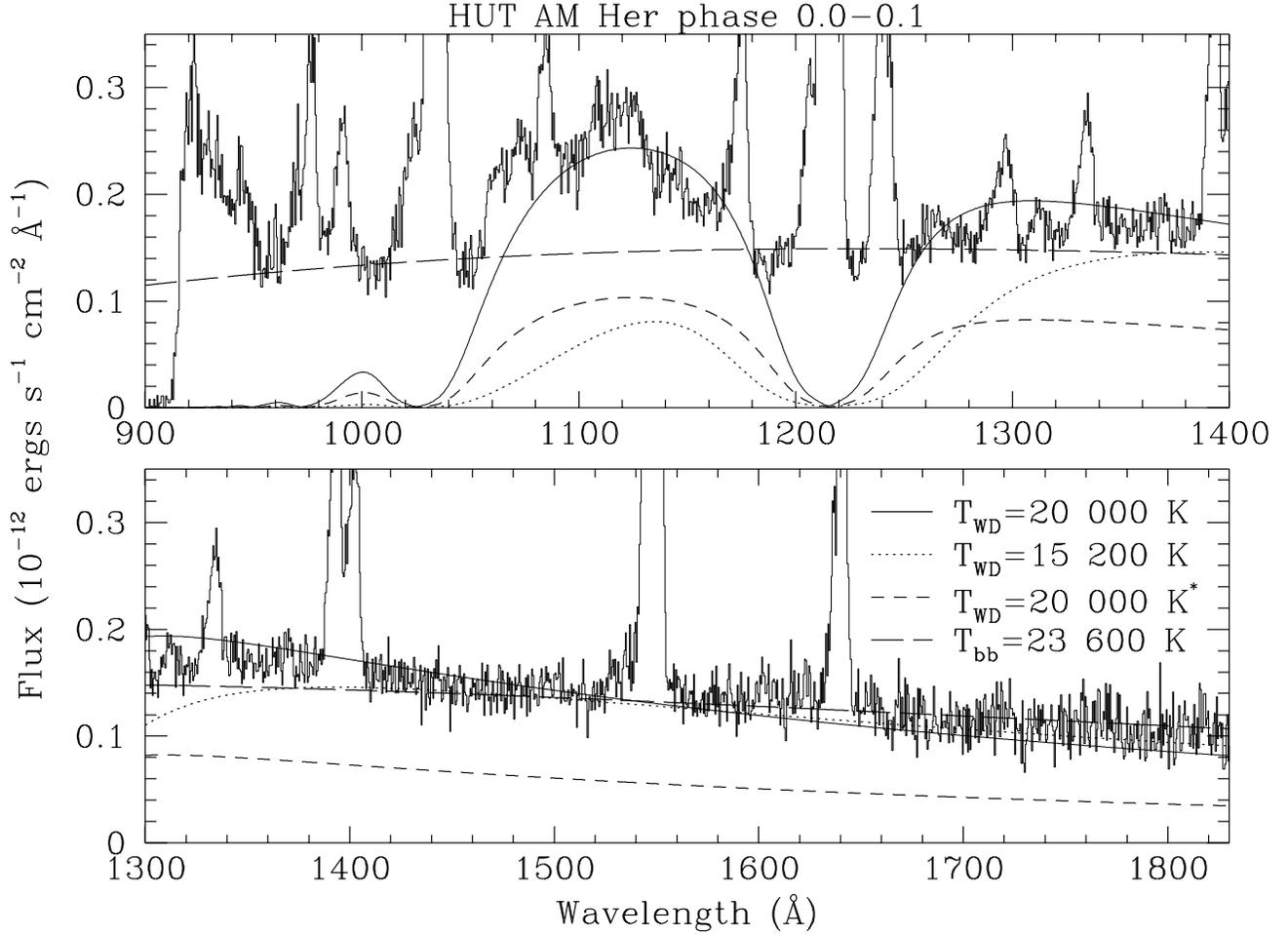}{6.5in}{270}{65}{65}{-254}{450}
\caption{\label{fig:allfit} 
Attempts to fit the faint phase continuum with white
dwarf models have not been successful.
White dwarf atmospheres are not able to fit the region below \LA. 
Three WD fits are shown: the best fit for a 20\thinspace000 K WD (solid line), 
the best WD fit (15\thinspace200 K) for wavelengths longer than 
1400 \AA\ (dotted line), 
and a 20\thinspace000 K WD normalized to the flux observed in the
low state with IUE (dashed line). 
The best fit is actually a blackbody fit with 
T=23\thinspace600 K, shown as the long dashed line.
}
\end{figure}

If the WD is the source of the faint
phase continuum emission,
we would expect it to be best fitted by a WD atmosphere model. 
In Figure~\ref{fig:allfit} we have attempted such a fit, without success,
for a grid of WD model atmosphere spectra constructed with 
the TLUSTY and SYNSPEC programs (Hubeny, Lanz, \& Jefferey 1995;
Hubeny \& Lanz 1995). The methods used to construct these models are
described in detail by Long \etal\ (1993).
The best fit WD is near 20\thinspace000 K,
but the quality of the fit is extremely poor.  The slope is too steep
at the longer wavelengths, and it misses the flux at the
shorter wavelengths completely. A cooler 15\thinspace000 K atmosphere
fits the spectrum longward of 1400 \AA, but fails to fit shorter wavelengths.
The basic problem is that the Lyman lines are absent in the observed
faint phase spectrum, but should be strong in a WD atmosphere at these 
temperatures.

The lack of absorption features in the AM Her WD spectrum has been noted
by Paerels \etal\ (1996), and Mauche \& Raymond (1998). 
The model of AM Her might lead us to expect this. If the shock occurs
at some height above the WD, producing hard X-rays, the atmosphere will 
be heated from the top down, and a WD atmosphere with such an 
inverted temperature structure would not produce Lyman absorption
(see van Teeseling \etal\ 1994). The most obvious problem with this, though, 
is that for the faint phase observation, little if any of the irradiated
portion of the WD should be visible. For the bright phase spectrum, if 
the accretion heats the photosphere directly, its structure should be
similar to a normal WD
atmosphere, and thus would expect
to see Lyman lines in the bright phase spectrum, as we do. 
We also might expect a contribution to the faint phase emission
from an uneclipsed portion of the hotspot, a second pole, or the
accretion column, and these might account for much of the UV emission.
Also plotted is a 20\thinspace000 K WD atmosphere normalized to the
flux observed in the low state with IUE (G\"ansicke \etal\ 1995,
Silber \etal\ 1996), demonstrating the change between the high and low states.

\begin{figure}
\plotfiddle{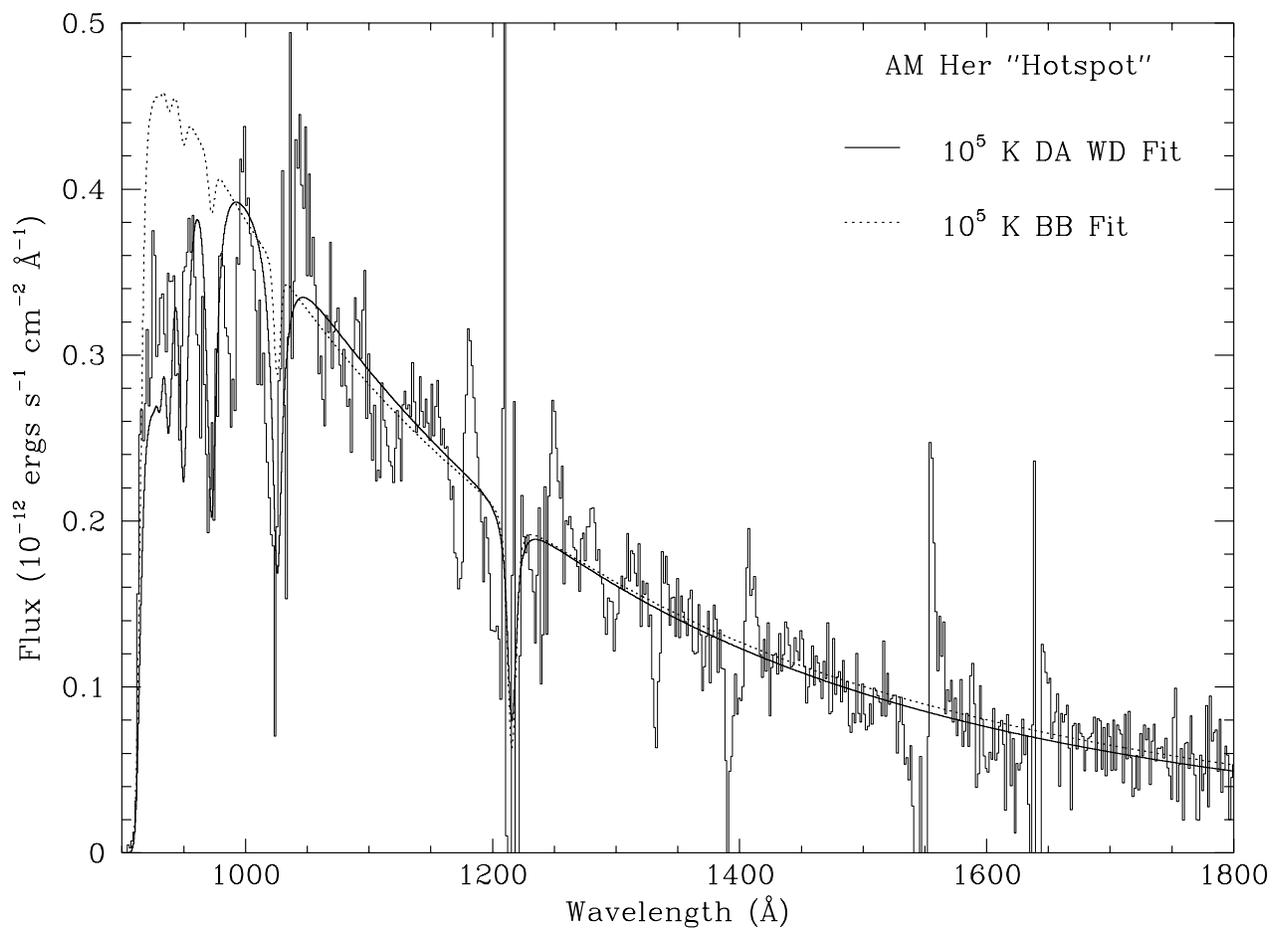}{6.5in}{270}{65}{65}{-254}{450}
\caption{ \label{fig:hotspot} 
The difference between the maximum and minimum spectra, shown here, is
assumed to be due to the hotspot. We have fitted 
100\thinspace000 K DA WD  and blackbody
spectra to it.  A neutral hydrogen column of $10^{20}$ 
cm$^{-2}$ was assumed in the blackbody fit.
Note that while many of the apparent absorption features are
caused by the subtraction of the changing line fluxes, the Lyman
line absorption is real. 
}
\end{figure}

Presumably the excess flux in the bright phase spectrum is seen because
the eclipsed portion of the hotspot on the surface of the WD 
has rotated into view. 
Making this
assumption, we subtracted the faint phase spectrum from the bright phase
spectrum to produce a hotspot spectrum, which is shown 
in Figure~\ref{fig:hotspot}, binned to 1.5 \AA.
Note that the subtraction
of lines whose positions are varying with phase creates a large
uncertainty near the line positions, and apparent absorption at these
wavelengths is not real. The Lyman line absorption, however, is real, as
can be readily seen by inspection of Figure~\ref{fig:plottwocont}.
We have fitted the unbinned spectrum
with a 100\thinspace000 K DA WD atmosphere. The fit suggests an emission
region of radius $2\times10^8$ cm, or about 2\% of the surface of the
WD.
The hotspot spectrum
shortward of 1000 \AA\ cannot be fitted by a blackbody.
A reasonable fit longward
of 1000 \AA, shown in Figure~\ref{fig:hotspot}  
with a neutral H absorption column of
10$^{20}$ cm$^{-2}$, 
greatly overpredicts the flux shortward of 1000 \AA.
Note that a 10$^{20}$ cm$^{-2}$ neutral H absorption column cannot
account for the depth of the Lyman lines we observe.

AM Her was observed with EUVE on the same day as the first two HUT
observations. The best fit to the peak EUVE spectrum is a 19 eV
blackbody with an area of $1.8\times10^{17}$ cm$^2$ 
(Mauche \etal\ 1998). The corresponding
flux at 1000 \AA\ is $3.5\times10^{-13} \OIGS$, very close to the value
we observe in the HUT hotspot spectrum. This suggests that the excess 
flux below \LA\ in the HUT wavelength range is the tail of the soft 
X-ray blackbody.  

\subsection{Emission Line Variations}

\begin{figure}
\plotfiddle{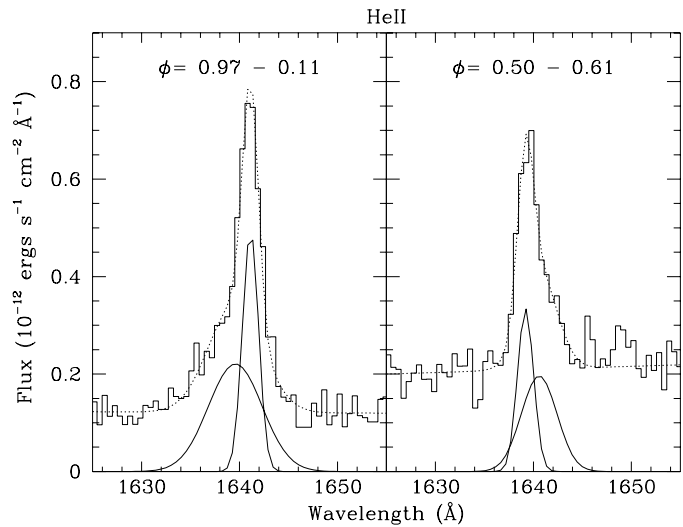}{6.5in}{270}{35}{35}{-140}{350}
\caption{\label{fig:plottwo}
The He II $\lambda1640$ line for faint and bright phases, respectively.
In each case the line has been fitted with two Gaussians. The width of the
narrow component has been fixed to HUT's instrumental resolution 
while the broad
width was allowed to vary, along with the line positions and fluxes.
For phase $0.97-0.11$, material in the accretion column is 
moving in our general
direction and the broad component appears blueshifted. At phase $0.50-0.61$,
the material in the accretion column is moving away from us and 
the broad component
appears redshifted.
}
\end{figure}

The emission line fluxes are lower during the bright phase
than during the faint phase.
Many of the emission lines appear to be comprised of  broad
and narrow components.
The \ion{He}{2}~$\lambda1640$ line is plotted in Figure~\ref{fig:plottwo}
for two phases. At phase $0.97-0.11$ a blueshifted broad component is
apparent, while the broad component is redshifted at phase $0.50-0.61$.
In an attempt to determine the relative amounts of emission arising
in the broad and narrow line regions, we have deblended the lines
using ``SPECFIT''. 
For the \ion{He}{2} line we modelled
the line with two Gaussians, whose width, position, and normalization
were allowed to vary to minimize $\chi^2$, 
with the constraint that the narrow line
width be equal to the HUT instrumental resolution at that wavelength. (HUT has
a wavelength dependent spectral resolution, 
see Kruk \etal\ 1995.) 
The assumption of Gaussian line profiles is an approximation. In reality
the narrow and broad components may be comprised of multiple components
themselves, which may arise in different parts of the system as 
evidenced by different radial velocity amplitudes observed for 
different lines (e.g. Crosa \etal\ 1981). 
ORFEUS observations
of the narrow line component of the \ion{O}{6}~$\lambda\lambda1032,1038$ 
lines (Raymond \etal\ 1995, Mauche \& Raymond 1998)
suggest that the narrow component is much less than the HUT resolution,
so the expected width should be the instrumental width. 
The resulting fits are shown in Figure~\ref{fig:plottwo}. 
At $\phi=0.97-0.11$ the broad component is blueshifted relative to 
the narrow component, while at $\phi=0.50-0.61$ it is redshifted.
Phase $0.97-0.11$ corresponds to part of the phase in which the
EUVE and Ginga observations indicate the hotspot is on the far side of
the WD (Paerels \etal\ 1996, Beardmore \& Osborne 1997), and so a 
blueshifted broad component 
is consistent with material moving in the accretion column toward
the far side of the WD, or toward us. 
At phase $0.50-0.61$, we 
then expect that material to be moving toward the near side of the
WD (away from us), consistent with the observed profile.

We performed similar fits for a number of lines on three subsets of the data. 
The first two of these include only data taken during orbital night, and are
shown in Figure~\ref{fig:plottwocont}. 
The third subset covers phases $0.67-0.81$
and includes data taken during orbital day.
To account for small shifts in the
positions of the lines on the detector because of pointing, we have
used the pointing errors 
reported in the telemetry to register and sum the spectra to produce
these data sets. 
For the \ion{C}{4}~$\lambda1548,1551$ doublet, the HUT resolution is
not sufficient to allow an unconstrained 4 Gaussian component fit. 
To determine the broad and
narrow contributions, we used four components, but constrained the doublet
components to be at a fixed separation,
have the same width (the instrument resolution for the narrow components), 
and have
the arbitrary, but reasonable, flux ratio of 1.3:1 (cf. \ion{Si}{4},
Table~\ref{tbl:fluxes}
and Raymond \etal\ 1995, Mauche \& Raymond 1998).
For the \ion{Si}{4}~$\lambda1394,1403$ doublet, only the separation
of the doublet components and the narrow component width were fixed. 
\ion{N}{5}~$\lambda1239,1243$ and \ion{O}{6}~$\lambda1032,1038$ 
were fit in the same manner as \ion{C}{4} (an airglow component was
added for the third observation \ion{O}{6} fit to account for 
terrestrial Ly $\beta$). 
For \ion{He}{2}~$\lambda1085$, the
widths and separation of the broad and narrow components were fixed to
the values obtained in the \ion{He}{2}~$\lambda1640$ fit (with the
widths corrected for the wavelength dependent HUT resolution). 
\ion{C}{2}~$\lambda1335$ and \ion{C}{3}~$\lambda1176$ were fit 
as \ion{He}{2}~$\lambda1640$ was, even though these lines are 
closely spaced multiplets. 
The remaining lines were fit in the same manner as \ion{He}{2}~$\lambda1640$.

In all cases, the broad component was blueshifted, relative to the narrow
component, at $\phi=0.97-0.11$, redshifted at $\phi=0.50-0.61$, 
and approximately coincident with the narrow line at $\phi=0.67-0.81$.
The broad and narrow fluxes determined are listed in Table~\ref{tbl:fluxes}.
The table also includes the total flux in the line determined by integrating
the flux above a linear fit to the continuum on either side of the line. 
The last row gives the total flux for each column.
The wavelength calibration combined with pointing uncertainties prevents
the determination of a velocity curve for our narrow line data. The
broad line velocities are consistent with the  
optical \ion{He}{2} velocity curves (Greenstein \etal\ 1977), with
large blueshifted velocities during the first observation ($\phi=0.97-0.11$),
large redshifted velocities during the second observation ($\phi=0.50-0.61$),  
and little velocity shift during the last observation ($\phi=0.67-0.81$).
In general, the fluxes for both components decrease across the observations. 
Most of the decrease in the narrow component flux occurs between
the second and third observations. This is consistent with 
the formation of the narrow
component on the face of the secondary, since
the optical velocity curves imply that the heated secondary face is 
most visible at $\phi= 0.16$ and most hidden at $\phi=0.66$.
This is not true for the \ion{N}{5} narrow component, 
which actually increases with
phase, or the \ion{O}{6} and \ion{C}{2} narrow components, 
which are approximately constant with phase. Portions of these components
may form in other parts of the system.
It is unlikely that the narrow component arises on the surface of the
WD. Zeeman splitting of the H Balmer $\alpha$ absorption line has been 
observed by Schmidt, Stockman, \& Margon (1981). We would expect
a similar splitting of the \ion{He}{2} Balmer $\alpha$ emission line
at 1640 \AA\ if this line was formed near the surface of the WD, and 
such a splitting is not observed. 

The velocity curves for the broad component imply that the portion of the
accretion stream producing the \ion{He}{2} line emission is most directed
toward us at $\phi=0.03$, 
and away at $\phi=0.53$. Since much of the line emission
does not appear to be 
optically thin (note our \ion{Si}{4} doublet ratio, and
the \ion{O}{6} doublet ratio in the ORFEUS data), we would expect to see the
most flux from the column when the the source of illumination is between
the accretion column and us. 
This is what we observe, as the maximum flux appears during the
first observation, while it is roughly constant for the other two. 

\begin{figure}
\plotfiddle{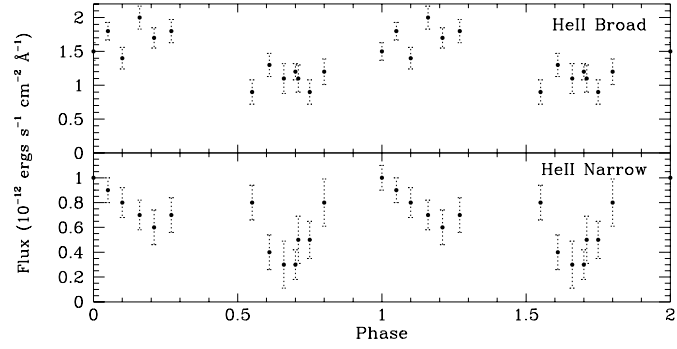}{3.5in}{270}{35}{35}{-140}{350}
\caption{\label{fig:Hedeblend}
The light curves for the He II broad and narrow
components. Note the partial eclipse of the narrow line
flux by the secondary  from 
phase $0.61-0.75$, centered on secondary conjunction at phase 0.66.
}
\end{figure}

Because the HUT resolution is best near the \ion{He}{2}~$\lambda1640$
line, and because it is not a multiplet, we were able to perform fits
for this line at higher temporal resolution. 
The resulting narrow line intensities are plotted in 
Figure~\ref{fig:Hedeblend} with phase. $\Delta\chi^2=1$ error bars are shown.
Between $\phi=0.55$ and 
$\phi=0.8$ the flux drops by more than a factor of 2. This drop is
centered on secondary conjunction ($\phi=0.66$, Greenstein \etal\ 1977),
so we must conclude that the \ion{He}{2} line region is partially 
eclipsed by the secondary star. Note that the data for 
$\phi=0.55-0.66$ and $\phi=0.67-0.80$ are from two different observations
separated by 4 days, making it unlikely that this is a transient effect.
A significant decrease in the narrow \ion{He}{2}~$\lambda4686$ 
narrow line flux is also
seen at $\phi=0.6$ in the optical data of Greenstein \etal\ (1977), although
those authors did not identify it with an eclipse.

\begin{figure}
\plotfiddle{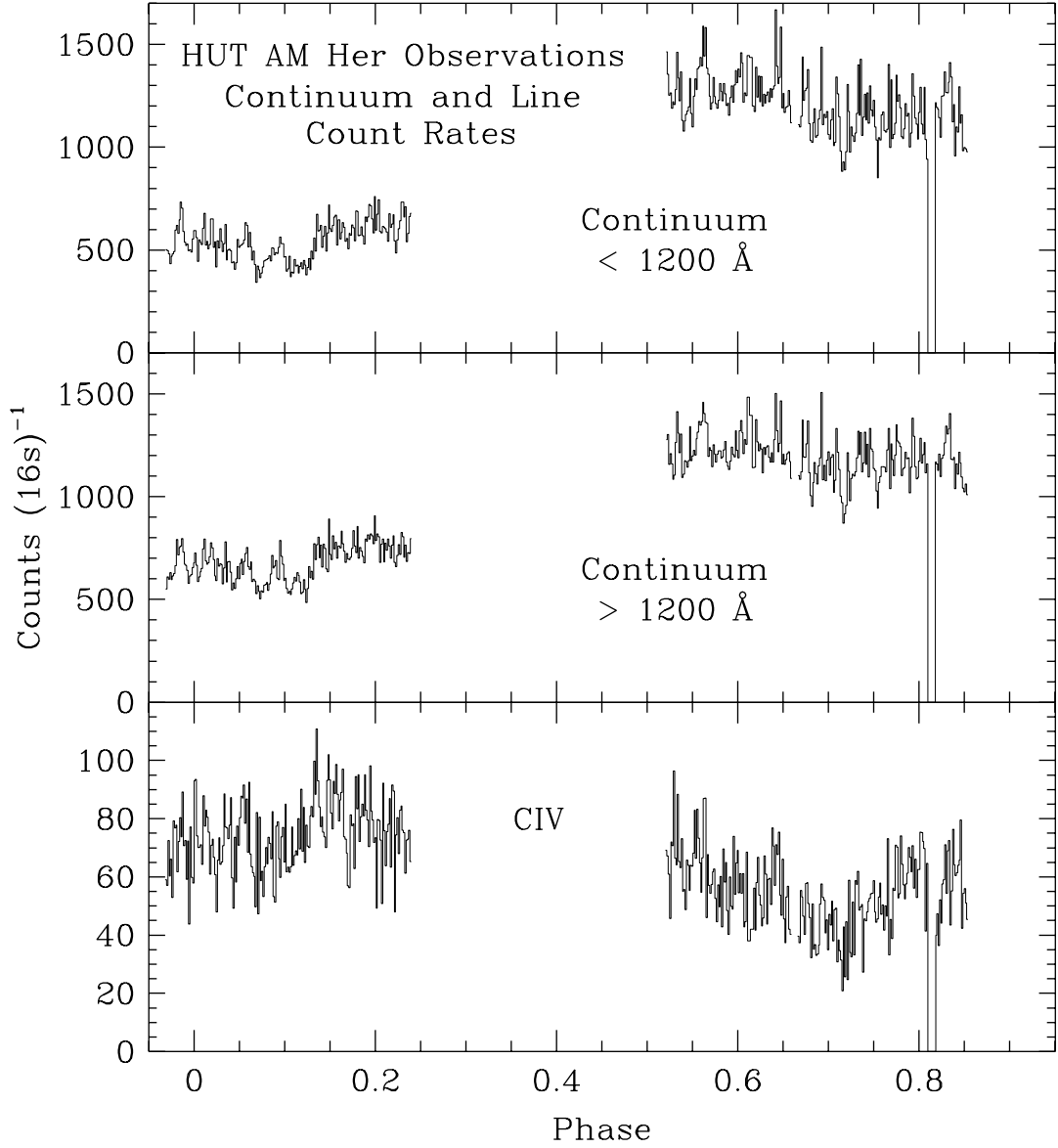}{6.5in}{0}{75}{75}{-240}{-100}
\caption{\label{fig:contvslines}
The count rate per 16 s for the line-free continuum regions and the 
continuum-subtracted C\thinspace IV line
are plotted vs. phase for the three observations. 
Note the larger change at short continuum wavelengths and the
general anticorrelation between the trends for line and continuum 
regions.
}
\end{figure}

We attempted to fit the \ion{C}{4}~$\lambda1549$ line as well. While
the fits were poor, they do suggest an eclipse in this line as well, 
and this is more clearly seen in the total line flux 
(see Figure~\ref{fig:contvslines}, discussed below; note that the total
line flux includes both the narrow and broad line components).
Unfortunately the low detector resolution at the \ion{O}{6} doublet position
(FWHM 1110 km s$^{-1}$ at 1035 \AA\ versus 350 km s$^{-1}$ at 
1640 \AA) combined 
with the presence of \LB\ airglow and \LB\ continuum absorption (for a total 
of 6 components to fit)
make it impossible to fit the \ion{O}{6} doublet with sufficient accuracy.
We note
that the Mauche \& Raymond (1998) observations of the \ion{O}{6} line do not
show a strong eclipse.
The presence of a narrow eclipse at secondary conjunction lends support to 
the claims of Davey \& Smith (1996) and Wickramasinghe \etal\ (1991) that
the inclination angle of AM Her is greater than $45\arcdeg$, rather
than the $35\arcdeg$ value normally assumed (Brainerd \& Lamb 1985). 

The \ion{He}{2} broad component is also plotted in the Figure. It shows
a decrease of about 35\% from phase $0.0-0.3$ to $0.5-0.8$. The phase dependence
is similar to \ion{O}{6} broad line light curve of Mauche \& Raymond (1998),
but much shallower than the 70\% decrease they observe. This suggests that
the \ion{O}{6} broad line flux is more dependent on viewing angle than is
the \ion{He}{2} broad line flux. 

\subsection{Rapid Variability}

\begin{figure}
\plotfiddle{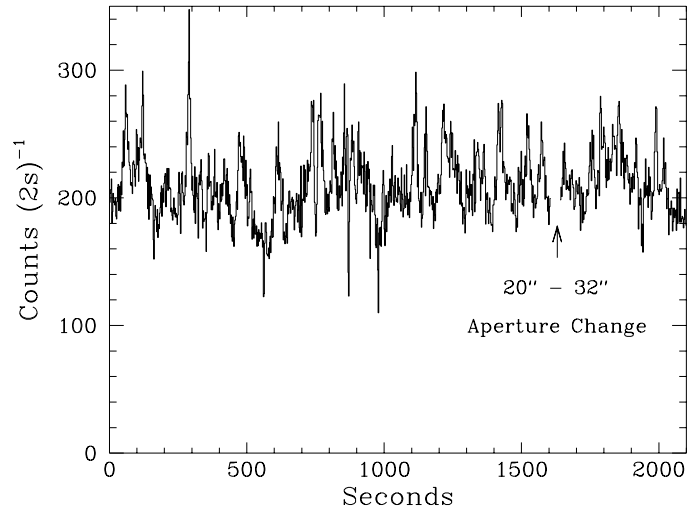}{6.5in}{270}{35}{35}{-140}{350}
\caption{\label{fig:3rdobstot} 
The total source count rate per 2 s is shown for the third observation.
The strong variability, with increases of more than 50\% in 10 s, is
intrinsic to AM Her. 
During the observation we changed the aperture from 20\arcsec\ 
to 32\arcsec\ to confirm that pointing and poor centering 
were not the sources of the 
variations.
}
\end{figure}

In addition to a general increase in the continuum
by a factor of two from faint to bright phase, rapid variability is seen
on timescales of $\sim$ 10 s.
The amplitude of these variations approaches 50\% of the total continuum
count rate in some cases.
The count rate variations were obvious in real-time during 
the observations, and are clear in the source count rate for 
the $\phi=0.67-0.81$ observation shown in  Figure~\ref{fig:3rdobstot}.
To rule out the possibility that the observed
variations were related to pointing or source miscentering in the aperture, 
we changed the aperture during
this observation from a 20\arcsec\ to a 32\arcsec\ circular aperture.
This had no effect on the mean source count rate and the 
variability continued. Also, as mentioned earlier, 
source centering was verified by post flight analysis of video frames
stored during the observation. Hence, we are confident this rapid 
variability is intrinsic to AM Her.

\begin{figure}
\plotfiddle{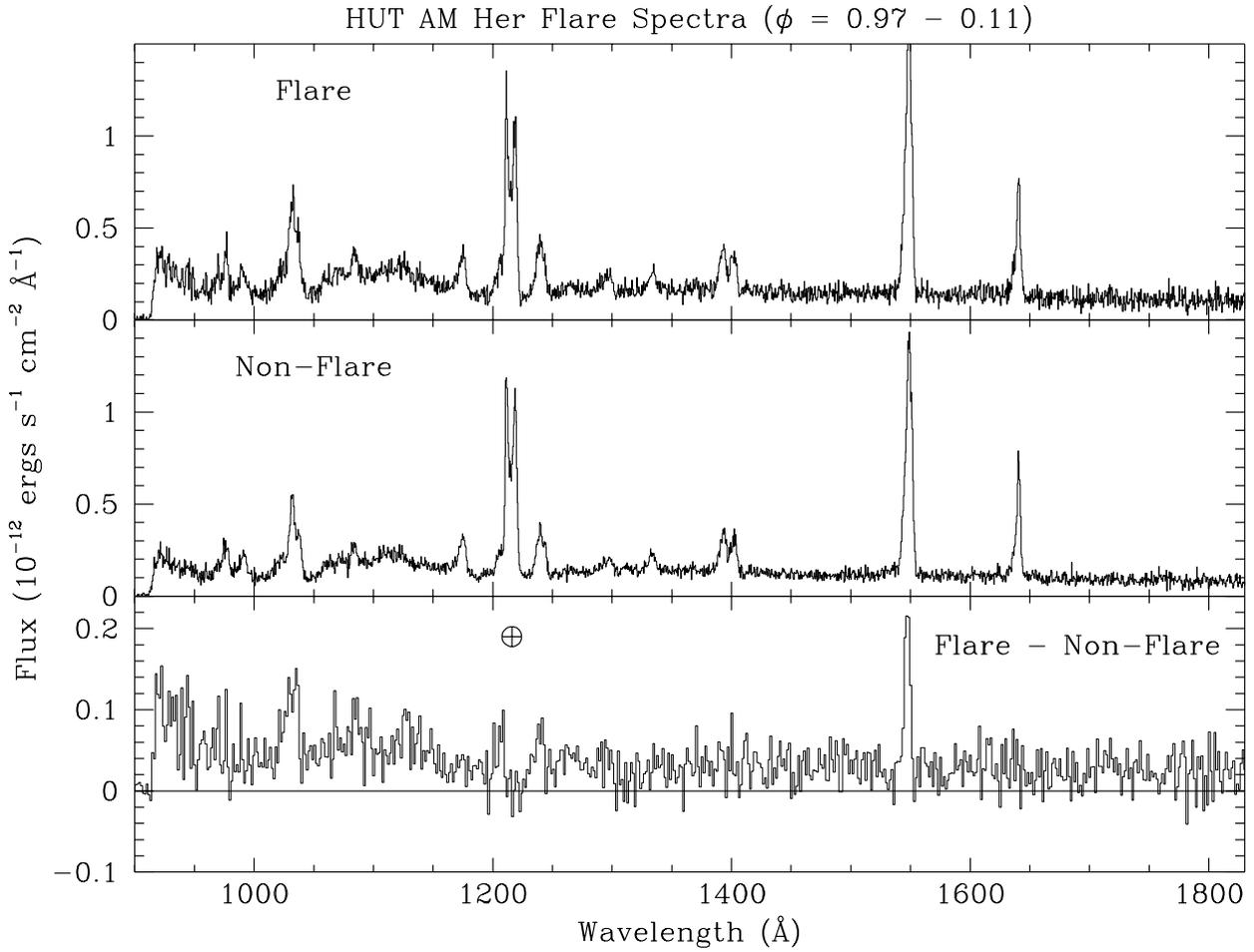}{6.5in}{270}{65}{65}{-254}{450}
\caption{\label{fig:flarespecobs1}
The ``Flare'' spectrum is the sum of all of the 
$\phi=0.97-0.11$ 2 s bins greater than
1/2 $\sigma$ above the local mean.
The ``Non-Flare'' spectrum is the sum of the bins
below the mean. The lower panel is the difference between the two.
Note the presence of line emission in the difference spectrum.
}
\end{figure}

\begin{figure}
\plotfiddle{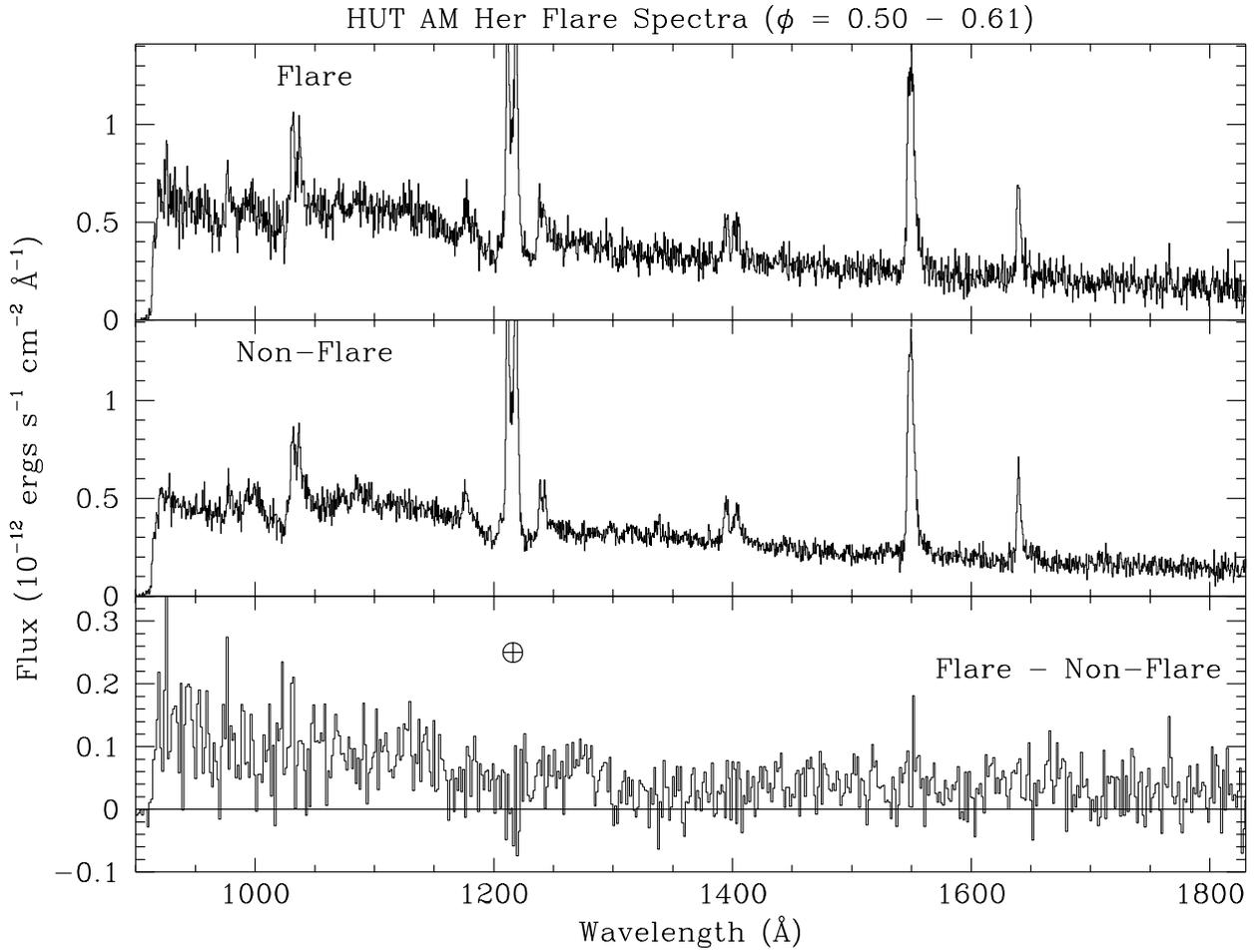}{6.5in}{270}{65}{65}{-254}{450}
\caption{\label{fig:flarespecobs2}
Same as Fig. 8, but for $\phi=0.50-0.61$. 
Note the absence of line emission in the difference spectrum.
}
\end{figure}

\begin{figure}
\plotfiddle{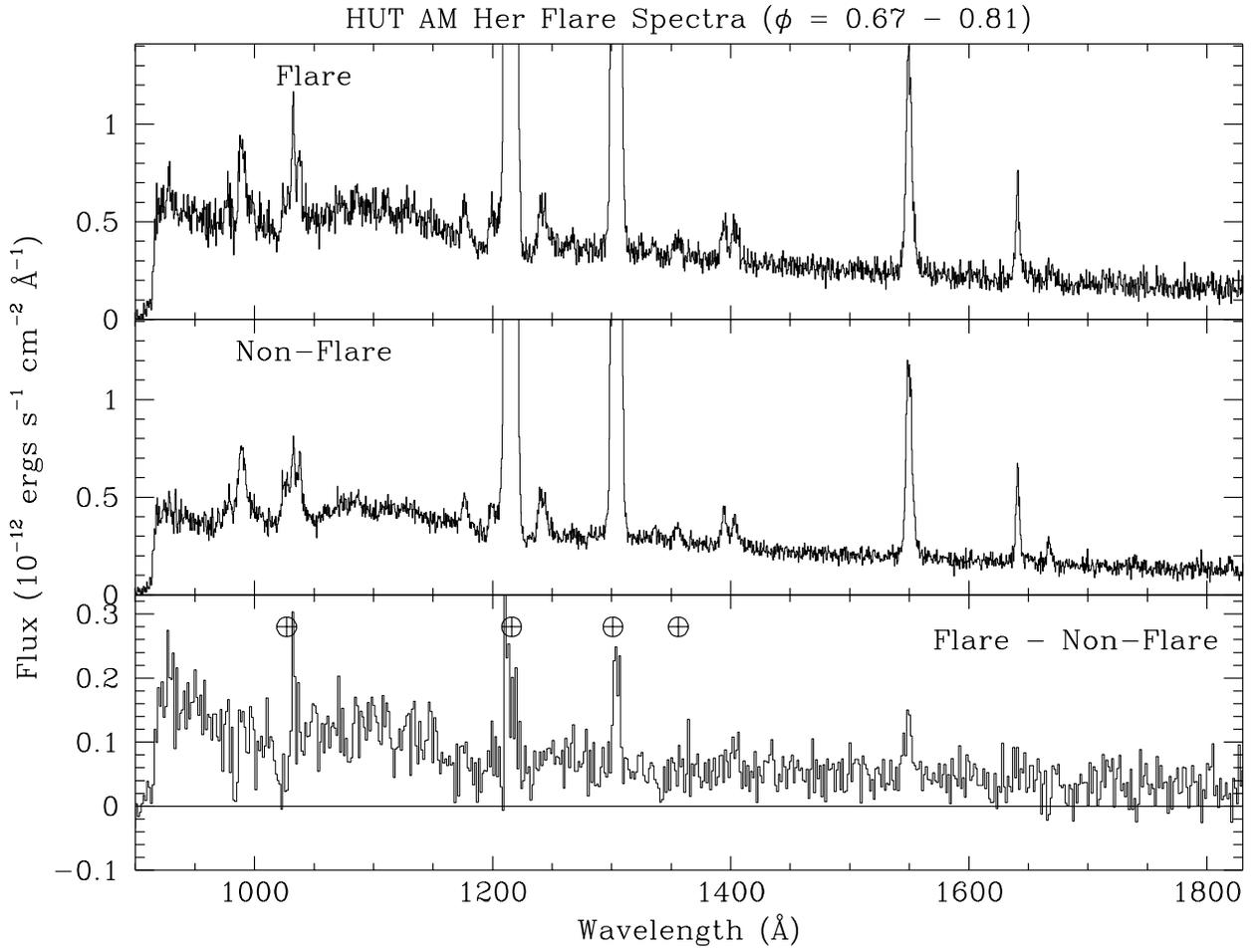}{6.5in}{270}{65}{65}{-254}{450}
\caption{\label{fig:flarespecobs3}
Same as Fig. 8, but for $\phi=0.67-0.81$.
}
\end{figure}

To investigate the nature of the flares, we have constructed their average
spectrum. For each observation, we first 
produced a spectrum (``Flare'') from all of the 
bins in each observation
with count rates more than 1/2 $\sigma$ above
a local mean (the local mean was defined by a fit to the count rate), 
and a spectrum (``Non-Flare'') from all of the bins below the 
mean. These are shown in 
Figures~\ref{fig:flarespecobs1}$-$\ref{fig:flarespecobs3} 
for each of the three HUT observations, along with 
the difference spectrum, which has been binned to 1.5 \AA. 
The difference spectrum for $\phi=0.97-0.11$, the faint phase,
shows narrow lines and a faint continuum. The difference spectra for
the other two observations, which were at bright phases,
have a relative excess of flux in the 912-1170 \AA\ region.
The $\phi=0.50-0.61$ difference spectrum does not have line 
emission, while the other observations do.
The magnitude of the airglow lines in the difference spectrum
is much less than the error. If there were a problem with our temporal
bin selection (if most of the bins came from a 
particular phase, for example)
we might expect to see measurable airglow lines, so their absence is 
reassuring.

Since we expect that the line emission arises in gas which reprocesses
the EUV and X-ray continuum emission,
we might expect to find some relationship between the variations 
of the line and continuum count rates. 
The line-free continuum and \ion{C}{4} count rates are plotted in
Figure~\ref{fig:contvslines}.
The continuum count rates are the sum of the regions 
$916-960$ and $1045-1160$ \AA,
and $1250-1280, 1410-1530, 1560-1620, 1650-1830$ \AA, respectively.
The light curves have been binned to 16 s. 
The low signal to noise ratio 
in the line count rate
makes it difficult
to see a correlation between the line and continuum light curves,
but the variation seen in the \ion{C}{4} light curve is not all noise.
A K-S test on the unbinned \ion{C}{4} data strongly 
implies that the variations are not Poissonian.  
We used Spearman's and Kendall's rank correlation tests on 
each observation, after the data were detrended with a fourth order
fit, to search for a correlation between the total continuum and
\ion{C}{4} light curves.
For $\phi=0.97-0.11$, Spearman's $r$ and Kendall's
$\tau$ both exclude the possibility that the two are uncorrelated at 
better than the 0.5\% significance level (meaning that if they were 
uncorrelated, they would produce $r$ and $\tau$ values larger than 
we calculated less than 0.5\% of the time). 
For $\phi=0.50-0.61$ the tests were inconclusive. For $\phi=0.67-0.81$,
they both excluded no correlation at the 1\% significance level. 
Thus overall it appears that there is a correlation between the line
and continuum flux variations.

The narrow line emission is strongest in the 
flare spectrum for the first observation,
when the hotspot is eclipsed, and the irradiated face of the 
secondary star most directly in view. There is almost no narrow 
line emission in the $\phi=0.50-0.61$ flare spectrum, which occurs near
secondary conjunction, when the largest fraction of the irradiated
face is out of view. Thus it is apparent that the line emission seen
in the $\phi=0.97-0.11$ flare spectrum is produced when the flare continuum
flux photoionizes material on the face of the secondary star.
The blue continuum in the bright phase flare spectra is most likely
from the hotspot on the WD. If it arose at the secondary star or in the 
accretion stream, we would expect it to be strongest in the 
$\phi=0.97-0.11$ observation
as the narrow and broad line emission is (cf. Figure~\ref{fig:Hedeblend}).

\begin{figure}
\plotfiddle{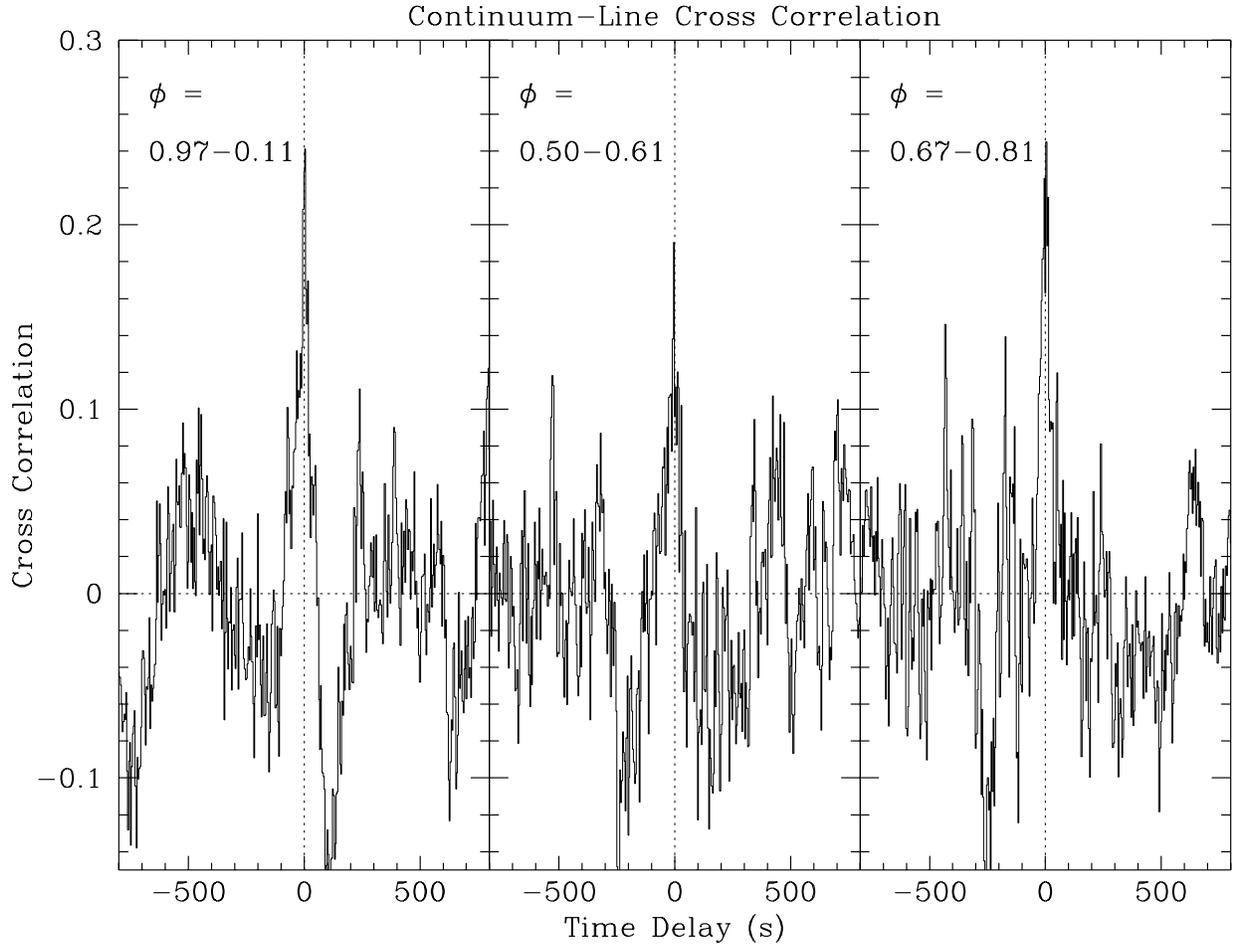}{6.5in}{270}{65}{65}{-254}{450}
\caption{\label{fig:ccfs}
The cross correlation functions for the continuum count rate and
C\thinspace IV line count rate for each observation. No evidence for
a time lag between the line and continuum count rates is seen.
}
\end{figure}

Observations of the polar ST LMi by Stockman \& Schmidt (1996) 
show a time lag between the line and continuum fluxes of about
$40-80$ s for that system. Szkody \& Margon (1980) show a weak 
correlation at a lag of 8 s between U band and 
\ion{He}{2}~$\lambda4686$ fluxes for AM Her. 
To investigate the possibility of a line-continuum 
time lag in our AM Her data we constructed the 
\ion{C}{4}-continuum cross correlation functions for each observation,
shown in Figure~\ref{fig:ccfs}. No evidence for a time lag 
is seen. 
The secondary star is less than two light-seconds from the WD 
and the ionization timescales are much less than this, so we would
not expect to detect the lag associated with X-rays from the WD 
ionizing material on the secondary star. 

The continuum autocorrelation functions for the three observations have 
e-folding times of 44, 18, and 28 s, respectively, suggesting that the typical
flare duration is of order of tens of seconds.
This is similar to flickering observed in the optical 
(Stockman \& Sargent 1979, Panek 1980).
The flux of the flare (difference) spectrum
in the HUT range ($912-1830$ \AA) is 6.4$\times10^{-11}$ 
erg~s$^{-1}$~cm$^{-2}$. If we assume this to be produced by a 
100\thinspace000 K  blackbody similar to the hotspot inferred above,
then the total luminosity of that blackbody would 
be $4.9\times10^{32}$ erg s$^{-1}$. Taking the length of a typical flare
to be 10 s,
the total energy of such an event is then $4.9\times10^{33}$ erg.
Assuming then that this energy is released by the impact of a
clump of material in the accretion stream onto the WD,
the implied mass for a typical clump is $3.6\times10^{16}$ g, or a 
temporary increase
of $3.6\times10^{15}$ g~s$^{-1}$ in the accretion flow.
This is about the same mass determined by Beardmore
\& Osborne (1997) from Ginga high time resolution hard X-ray observations.
If instead we assume that the flare continuum we observe is the tail
of the 200\thinspace000 K blackbody observed in the EUV, 
then a similar calculation gives a clump energy
of $2.0\times10^{34}$ erg and a mass rate of $2.0\times10^{16}$ g~s$^{-1}$.
If we are to spread this mass over the extent of the hotspot, which from both our
observations, and the EUVE observations (Mauche \etal\ 1998) is 
found to be an area of about $1\times10^{17}$ cm$^{-2}$, the average density
is very low. Even using the higher mass rate derived above, it is just
$2\times10^{14}$ cm$^{-3}$. 

Frank, King, \& Lasota (1988) derive the minimum 
density for the accreting material to not encounter a shock before 
reaching the WD photosphere to be about $1\times10^{17}$ cm$^{-3}$. 
If we assume, as suggested by Kuijpers \& Pringle (1982), that the
soft X-ray excess in AM Her is produced by blobs of material imparting
their energy directly to the photosphere of the WD, heating the hotspot
to soft X-ray temperatures, then the filling factor must be less than 
0.002 for the blobs to be dense enough. 
The cooling time for such blobs is extremely short, so that this
low filling factor implies that only 0.2\% of the hotspot region is
emitting at any one time.  This is inconsistent with the emitting
area required to match the observed UV brightness. Therefore,
the Kuipers \& Pringle (1982) clumpy accretion picture proposed to
account for the excess soft X-ray emission of AM Her cannot account
for the rapid variability seen in the FUV continuum of AM Her. 
This casts serious doubt on the viability of clumpy accretion as an
explanation for the soft X-ray excess. Instead, a heating mechanism
capable of providing modest energy deposition over a large area is
required, such as the quasi-steady nuclear burning suggested 
for AM Her by Fabbiano \etal\ (1981) and more recently for 
supersoft sources (see van Teeseling 1998 for a review). 
Several authors have computed nuclear burning rates assuming spherically 
symmetric accretion onto a WD, (e.g. Katz 1977; Paczy\'{n}ski \& 
\.{Z}ytkow 1978; Weast \etal\ 1979; Iben 1982) 
and find that stable nuclear burning
is possible for accretion rates greater than about 10$^{18}$ g s$^{-1}$.
If instead of spherical accretion, the accreted matter is magnetically 
confined to a small region of the WD surface, as in AM Her, this scales to 
about 10$^{16}$ g s$^{-1}$ assuming the UV hotspot size we derive, 
or about $5\times10^{14}$ g s$^{-1}$ assuming
the EUV hotspot size derived by Sirk \& Howell (1998). 
The best fits for quasi-simultaneous observations with ASCA (see below) and
EUVE (Mauche \etal\ 1998) suggest peak luminosites of 
$10^{32.3}$ and $10^{34.3}$ erg s$^{-1}$ for the bremsstrahlung
and blackbody components, respectively. 
The bremsstrahlung luminosity implies a lower limit to the mass
accretion rate of $1.5\times10^{15}$ g s$^{-1}$, about 
the rate required for stable burning. If this mass
undergoes nuclear burning, the energy released is $10^{34.0}$ 
erg s$^{-1}$, consistent with the EUVE observation. 

\section{Modelling the Narrow UV Line Emission}

AM Her was observed with 
EUVE (PI C. Mauche) and ASCA (PI M. Ishida) on the same day the
HUT observations were made.  
Mauche \etal\ (1998) have modelled the EUVE data at the peak phase
in terms of a blackbody spectrum with 
$L=10^{34.2}$ erg s$^{-1}$, $T=10^{5.34}$ K, 
extincted by a neutral hydrogen column of N$_H = 9\times10^{19}$ cm$^{-2}$.
A detailed analysis of the ASCA data has not yet appeared. Therefore
we have retrieved the data from the HEASARC archive, and were able
to fit the high phase spectrum in terms of a bremsstrahlung
spectrum with $L=10^{32.3}$ erg s$^{-1}$,
$T=10^{8.5}$ K, 68\% covered by a neutral hydrogen
column of $10^{23}$ cm$^{-2}$.
Assuming this ionizing continuum, we have attempted to model
reprocessing on the surface of the secondary, which should produce 
some of the narrow emission line components.
Since we were interested in the lines from highly ionized gas, we have
only modelled the outermost part of the atmosphere, which by stellar
physics standards is tenuous, with densities $\lesssim~10^{14}$ cm$^{-3}$. 
We thus were
able to use the CLOUDY 90.04 code (Ferland \etal\ 1998)
for the modelling.

We first divided the irradiated secondary Roche lobe into 9 annuli,
which were subdivided azimuthally into 100 gridpoints,
with appropriate viewing angles and distances from the WD calculated
assuming the parameters in Table~\ref{tbl:param} and the equations for the
Roche potential.
Next we calculated 
the effective gravity at each grid point, including the gravities of 
the secondary and WD 
as well as the orbital acceleration, 
and then computed the pressure as a function of depth
assuming simple hydrostatic equilibrium and constant temperature.
Self consistency between the temperature used in this scale height
calculation and the temperature in the line forming
region as calculated by CLOUDY for a given distribution was obtained with
a temperature of about 25\thinspace000 K for all of the models. 
(Note the temperature was fixed only to compute an approximate
pressure structure. It was not fixed in the CLOUDY models). 
We imposed this pressure structure on a CLOUDY model irradiated
by a  log $L=34.2$ erg s$^{-1}$,
log $T=5.34$ K blackbody and a  log $L=32.0$ erg s$^{-1}$,
log $T=8.5$ K bremsstrahlung spectrum, multiplied by
the cosine of the incidence angle
at each gridpoint, and
multiplied by 0.8 to account for the angle between the direction to
the secondary and the position of the hotspot on the WD 
(assuming the magnetic colatitude and azimuth given in
Table~\ref{tbl:param}; note that the hotspot may not actually
be coincident with the magnetic pole). 
CLOUDY divides each gridpoint model into zones, such that
the relative ionization of the important elements does not
change by more than 0.15 over a zone, and then balances the
heating and cooling for the zone, working its way through the model.
For each attempt to obtain this balance, we impose the pressure we
have computed for that depth in the atmosphere. Once CLOUDY has 
traversed the entire model, which we
stop when $n_e \ll n_H$ (this occurs at a density
of $10^{13}-10^{14}$ cm$^{-3}$, depending on the temperature
and density structures for a given gridpoint), 
it repeats the process until
the H~$\alpha$ optical depth scale does not change by more
than 10\% on the next to last iteration. 
The line luminosities predicted for each grid point were then summed,
after correcting for the viewing angle. 

The predicted values for the phase
at which the irradiated face is most directly
facing the observer are listed in 
Table~\ref{tbl:NLresults}. The flux relative to \ion{C}{4} is listed
as well as the actual \ion{C}{4} flux.
The values for the narrow component luminosity 
as observed with HUT and IUE (Raymond \etal\ 1979) 
at this phase are also listed.
The model is a partial success. It predicts significant emission
from all of the observed species, and does not predict any other significant
line emission in the HUT wavelength range.
Most of the predicted  line ratios agree with the observed ratios.
All are within 40\% of the observed value,
with the exception of the highest ionization lines
\ion{N}{5} and \ion{O}{6}, which are underpredicted
(and \ion{N}{3}~$\lambda991$, but this is likely due to the poor 
quality of the fit to the observation, see Table~\ref{tbl:fluxes}).
The predicted \ion{C}{3}~$\lambda1909$ flux is very low, consistent with the
lack of a detection of this line with IUE (Raymond \etal\ 1979).
The predicted \ion{C}{3}~$\lambda977/\lambda1176$ ratio of 1.3 is 
higher than the observed 0.7. We note though, that
although the extinction for AM Her is low, E$_{B-V} < 0.05$, (Raymond \etal\
1979), it may affect the observed flux at these short wavelengths. 
If E$_{B-V}$ were as high as 0.05, the \ion{C}{3}~$\lambda977$ flux would need
to be corrected by a factor 2.2, 
the \ion{C}{3}~$\lambda1176$ flux by
a factor of 1.7, and the \ion{C}{4}~$\lambda1549$ flux by 1.6
(assuming mean galactic extinction, 
as described by Cardelli, Clayton, \& Mathis 1989).  

The model shows good agreement in the optical depth of the lines. The
doublet ratios are close to, but not quite, 1:1, as observed with our
\ion{Si}{4} doublet and the ORFEUS observations of the \ion{O}{6} doublet
(Mauche \& Raymond 1998, Raymond \etal\ 1995).  
The \ion{O}{6} line arises in gas with densities between about $10^{12}-
10^{13}$ cm$^{-3}$, while most of the other lines arise in gas of about
$10^{13}$ cm$^{-3}$. While the fractional ionization of \ion{O}{6} is
highest between $10^{10}-10^{11}$ cm$^{-3}$, the
atmosphere's exponential increase 
in density with depth is more important in determining the line emissivity. 
Thus the density at which the line forms is basically determined by the  
density near the point in the atmosphere where the ionizing photons 
for a particular species are exhausted. 

The predicted fluxes are about a factor of 3 too high, but
this is not surprising because  the model does not take into account
attenuation by the accretion stream,
both with respect to the illumination of the secondary and our view of the
secondary. \ion{Na}{1} doublet maps of the secondary (Davey \& Smith, 1996;
Southwell \etal\ 1995) show that the secondary is not evenly irradiated,
and this is likely to be a significant source of error in our model.

Note that our model has not been adjusted
to attempt to fit the HUT data, we have only input the EUVE and
ASCA data into our model of the narrow line region. 
The only free parameter is the turbulence, which we have given 
a value of 15 km s$^{-1}$.  
Increasing the turbulence increases the luminosity
of the lines overall (by about a 
factor of 2 for a turbulence of 80 km s$^{-1}$),
while it decreases the flux of the \ion{He}{2} lines
relative to \ion{C}{4}.
We have run a number of models to
explore the sensitivity of different lines to the input parameters.
Table~\ref{tbl:NLresults} gives the results for a model with half
the blackbody luminosity as our nominal model (halving the 
bremsstrahlung luminosity has no effect on the model). 
Note that though the luminosity changes by a factor of 2 from
the low luminosity model to the nominal model, the 
\ion{He}{2}~$\lambda1640$ flux only increases by a factor of 1.6, as
line saturation becomes important.  This casts doubt on the use
of the \ion{He}{2} line as an indirect indicator of the EUV 
luminosity, e.g. Patterson \& Raymond (1985). 
The \ion{O}{6} flux is more sensitive to the incident luminosity,
as it changes by a factor of 2.7 when the luminosity is doubled,
and thus may be more useful as a diagnostic of the EUV spectrum.
The table also lists the results of the low luminosity model with
a blackbody temperature of 25 eV. The \ion{O}{6} flux increases by
a factor of more than 2, while the \ion{C}{4} flux only changes modestly.

\section{Conclusions}
We observed AM Her with the Hopkins Ultraviolet Telescope, producing 
high time resolution FUV spectra of this object
including the region from $912-1200$ \AA. 
We discovered that the continuum emission, which longward of \LA\ has 
previously been described well with simple models, has significant
emission below \LA, especially 
in the $1050-1170$ \AA\ range and near the Lyman limit.
The continuum near the Lyman limit rises by more than a factor of 
two from the faint to the bright phase, and  
WD models are unable to fit the faint phase spectra. 
The hotspot spectum, taken as the difference of the bright and faint phase
spectra, is well fit by a 100\thinspace000 K WD model with the spot covering
about 2\% of the WD surface.

We observed the 
broad line components to shift with phase consistent with their formation
in the accretion column. We have fit the narrow and broad line components
separating the emission from different sites in the system and 
yielding phase resolved measurements of the UV line emission from
these sites. A partial eclipse of the \ion{He}{2} narrow line emission occurs
at secondary conjunction, implying that the inclination of the system is
$\gtrsim 45\arcdeg$.

AM Her is strongly variable on short timescales in the FUV. We observed
rapid variations of more than 50\% on timescales of $\sim$ 10 seconds. 
Analysis of the ``Flared'' portion of the data suggests that when the
irradiated face of the secondary is in view, the line and continuum
flares are correlated, and we see narrow line emission in the flare
spectrum at these times.  A blue continuum is seen in the flare
spectrum when the hotspot is in view.
Thus we see a direct response from the material on the face of the
secondary to changes in continuum flux on the WD. 
The rapid variability is not consistent with the clumpy accretion
model proposed by Kuipers \& Pringle (1982) to explain the 
soft X-ray excess in polars. Our observations are more
consistent with the nuclear burning suggested by Fabbiano \etal\ (1981),
and we find that given a mass accretion rate derived from our
fits to the ASCA bremsstrahlung emission, the rate energy is liberated by
steady nuclear burning is consistent with a simultaneous observation
of the EUVE blackbody luminosity.

We have constructed a model of the irradiated surface of the secondary, 
using EUVE and ASCA observations to fix the irradiating spectrum. 
The line emission predicted by our model of the irradiated secondary
is in reasonable agreement with our narrow line
observations for all but the highest ionization lines.

\acknowledgments

We would like to thank everyone who helped make the Astro-2 mission a
success, including the mission support personnel at NASA, the crew of
{\em Endeavour}, and our colleagues on the HUT team. We thank M. Ishida
and C. Mauche for coordinating simultaneous ASCA and EUVE observations,
and C. Mauche for useful comments and fits to the EUVE spectra.
Support for this work is provided by NASA contract  NAS5$-$27000 to 
Johns Hopkins University.

%%%%%%%%%%%%%  REFERENCES  %%%%%%%%%%%%%%%%%%%%%%%%%%%%%%%%%%%%%

%%%%%%%%%%%%%  FIGURES  %%%%%%%%%%%%%%%%%%%%%%%%%%%%%%%%%%%%%

%%%%%%%%%%%%%  TABLES  %%%%%%%%%%%%%%%%%%%%%%%%%%%%%%%%%%%%%

\clearpage
\begin{table}
\plotfiddle{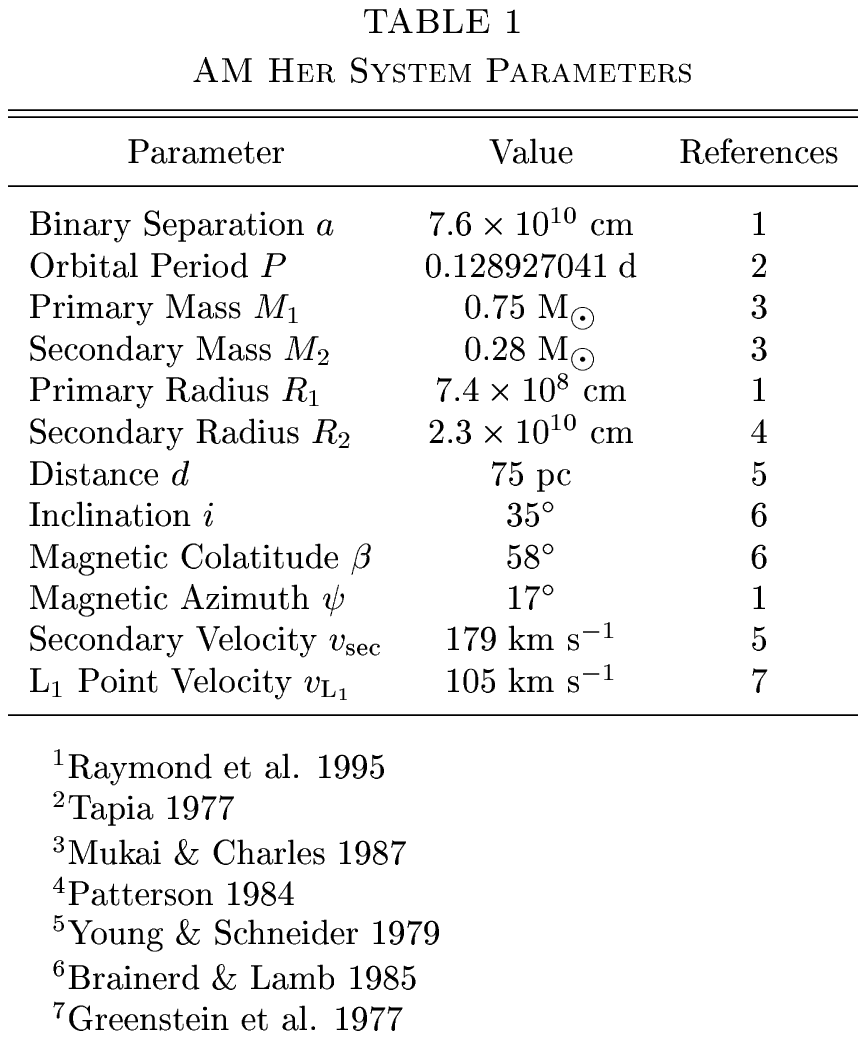}{7in}{0}{100}{100}{-300}{-200}
\dummytable
\label{tbl:param}
\end{table}
\clearpage

\pagebreak
\begin{table}
\plotfiddle{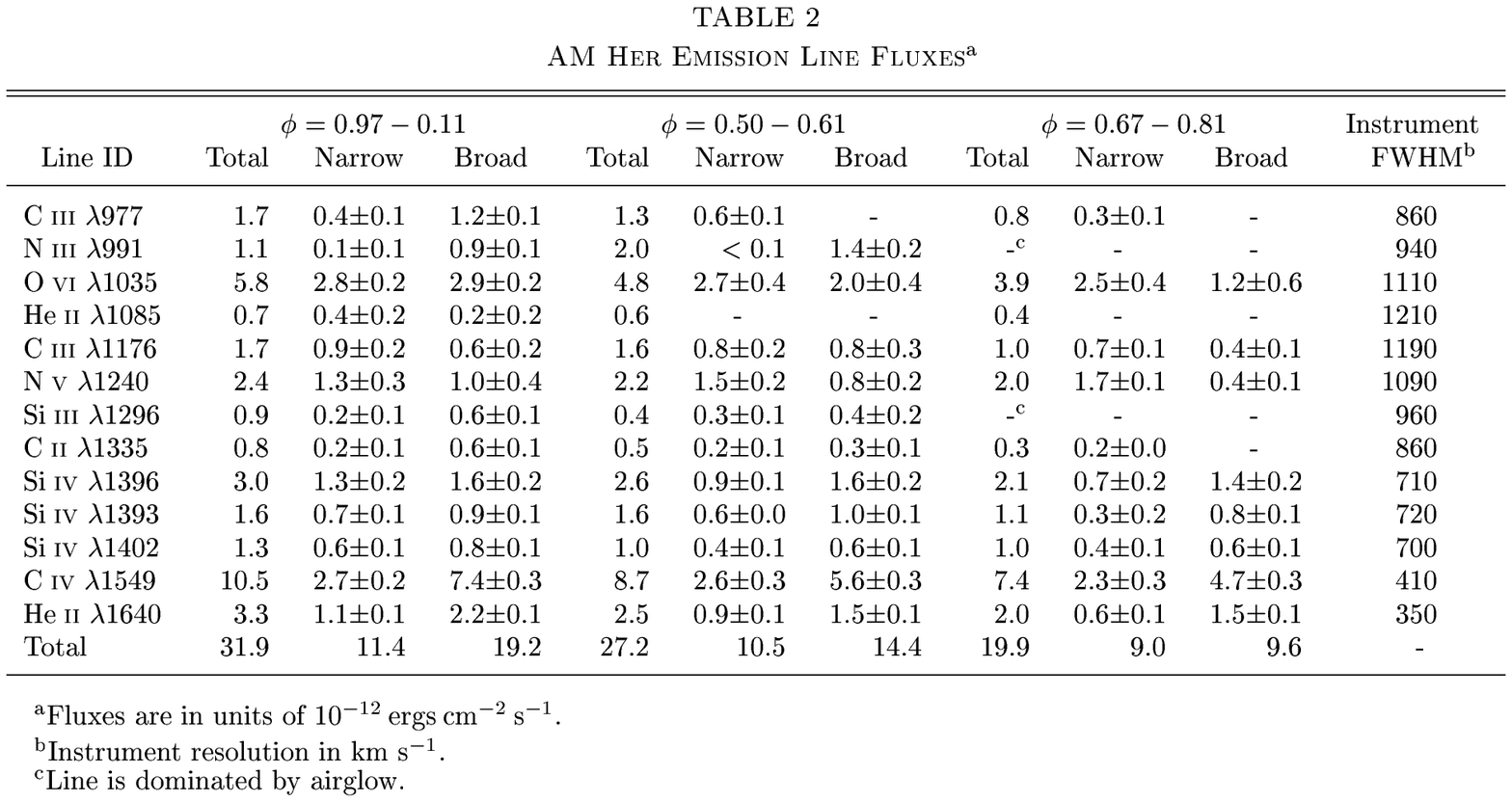}{7in}{0}{100}{100}{-300}{-200}
\dummytable
\label{tbl:fluxes}
\end{table}
\clearpage

\begin{table}
\plotfiddle{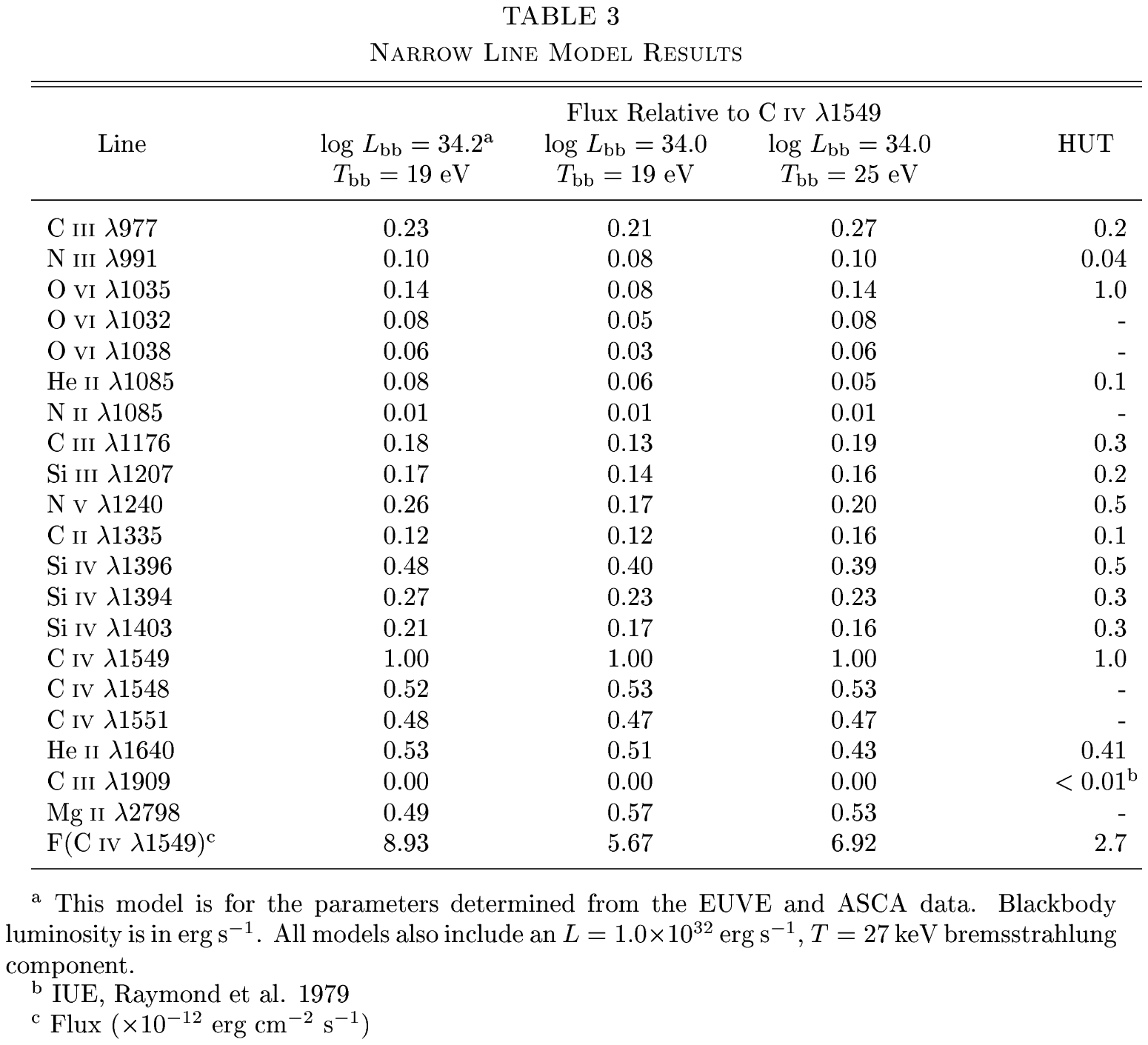}{7in}{0}{100}{100}{-300}{-200}
\dummytable
\label{tbl:NLresults}
\end{table}

\end{document}